\documentstyle[preprint,eqsecnum,aps]{revtex}
\begin{document}

\tightenlines
\draft
\preprint{SNUTP 99-005}
\title{Critical behavior in the rotating D-branes}

\author{Rong-Gen Cai }
\address{Center for Theoretical Physics, Seoul National
       University, Seoul 151-742, Korea}
\author{Kwang-Sup Soh}
\address{Department of Physics Education, Seoul National University,
   Seoul 151-742, Korea}
\maketitle

\begin{abstract}
The low energy excitation of the rotating D3-branes is thermodynamically
stable up to a critical angular momentum density. This indicates that there
is a corresponding  phase transition of the ${\cal N}$=4 large $N$
super Yang-Mills theory at  finite temperature. On the side of supergravity,
we investigate the phase transition in the grand canonical ensemble 
and canonical ensemble.  Some critical exponents of thermodynamic quantities
 are calculated. They obey the static scaling laws.  Using the scaling laws 
related to the correlation length, we get the critical exponents of the
 correlation function of gauge field. The thermodynamic stability of  
low energy excitations of the rotating M5-branes and rotating M2-branes 
is  also studied and similar critical behavior is observed.  We find that
the critical point is shifted in the different ensembles and there is no
critical point in the canonical ensemble for the rotating M2-branes. We also 
discuss the Hawking-Page transition for these rotating branes. In the grand
canonical ensemble, the Hawking-Page transition does not occur. In the 
canonical ensemble, however, the Hawking-Page transition may appear for 
the rotating D3- and M5-branes, but not for the rotating M2-branes.

\end{abstract}

\section{Introduction}

 Maldacena \cite{Mald} has  conjectured that there is a duality 
between the large $N$ limit of certain conformal field theories in various 
dimensions and supergravity on the product of anti-de Sitter spacetimes, 
spheres and other compact manifolds. Thus, some questions concerning
 large $N$ gauge theories may  be answered via supergravity. For instance, 
 one can calculate the correlation functions of the gauge 
theory \cite{Gubser1,Witten1}, study  thermodynamics and phase 
structure of the strong coupling gauge theory 
\cite{Itzhaki,Witten2,Li1,Barbon,Li2,Martinec,Gao}.

According to the Maldacena's conjecture,  thermodynamics of  certain 
conformal field theories in  the large $N$ limit can be described by 
thermodynamics of black holes in anti-de Sitter spacetimes. And the 
latter is found to be 
 quite different from thermodynamics of black holes in flat spacetimes.
 Hawking and Page \cite{Hawking} found that the heat capacity  is positive 
for large mass black holes, while is negative for small mass black holes. 
This indicates that there is a phase transition for the corresponding 
conformal field theory. In the high temperature phase, the black hole 
dominates the contribution to the conformal field theory; in the low 
temperature phase, the thermal anti-de Sitter space makes the dominant 
contribution. This phase transition has been interpreted as the
confinement/deconfinement transition in the Super Yang-Mills
 theory \cite{Witten2}.

As an explicit example of the  Maldacena's conjecture, the low
energy fluctuations of $N$ coincident D3-branes are supposed to be 
described by ${\cal N}$=4 super Yang-Mills theory with gauge group $U(N)$ 
in four dimensions \cite{Witten3}. Indeed, the entropy of the near extremal
black three-branes in the type IIB supergravity can be correctly reproduced
by the super-Yang-Mills theory \cite{Gubser2}, apart from a numerical 
factor. With this in mind that the result in supergravity should correspond 
to that for the strong coupling gauge theory and the result in \cite{Gubser2}
was given for the free gauge theory, the disagreement therefore is not a 
matter.  More recently, the coupling constant dependence in the 
thermodynamics of ${\cal N}$=4 super Yang-Mills theory has been 
investigated \cite{Gubser3}.  Considered the strong coupling limit, 
the result of the super Yang-Mills theory is supposed to be in 
agreement with that derived  from supergravity.

In the past few months, much attention has been focused on  the  rotating 
D-branes. For instance, the rotating D-branes has been used to 
extend Witten's QCD model \cite{Witten2} by introducing angular momentum
parameters in order to decouple the unwanted Kaluza-Klein particles 
\cite{Russo,Csaki}. At zero temperature the Coulomb Branch of ${\cal N}$=4 
super Yang-Mills theory is described in supergravity by multi-center 
solutions with D3-brane charge. At finite temperature and chemical potential 
the vacuum degeneracy is lifted, and minima of the free energy are shown
 to have a supergravity description as rotating black D3-branes \cite{Kraus}. 
In addition, Gubser \cite{Gubser4} has found that the spinning black 
three-branes in the type IIB supergravity are thermodynamically stable up to
a critical value of the angular momentum density. Inside the region of 
thermodynamic stability, the free energy from supergravity can be roughly 
reproduced by a naive model based on the free ${\cal N}$=4 super Yang-Mills 
theory on the world volume. The field theory model  can correctly
 predict a limit on the angular momentum density. Recall that the angular
momentum density corresponds to the density of R-charge of gauge field 
on the world  volume. This implies that there exists a phase transition 
of the ${\cal N}$=4 super Yang-Mills theory at the critical value of 
the R-charge density.

In the present paper we will study the phase transition of the super
 Yang-Mills theory on the side of supergravity by
 looking at the critical behavior 
in the grand canonical ensemble and canonical ensemble, respectively.
 Some critical exponents related to thermodynamic quantities are obtained 
and they satisfy the static scaling laws. Further we derive the critical 
exponents  of the correlation function of gauge field. We also discuss
the Hawking-Page transition in the rotating black branes. 
The critical behavior in the rotating D3-branes will be discussed
in  the next section. In sections III  and IV we extend this discussion 
to rotating M5-branes and M2-branes and investigate this kind of 
phase transition for the corresponding conformal fields on the world volumes,
respectively. We present our main results in section V.

\section{Critical behavior  in rotating D3-branes}

\subsection{ The solution  and thermodynamics}

The  rotating black three-branes in the type IIB supergravity may have three
angular momentum parameters. Since we are interested in the critical behavior
occurring at the critical angular momentum density, to show the scaling 
behavior, it is sufficient to  consider an angular momentum parameter 
nonvanishing only.  Such kind of solution has been given in 
\cite{Russo,Gubser4,Youm}. The metric is
\begin{eqnarray}
\label{e1}
ds^2 &=& \frac{1}{\sqrt{f}}\left(-h dt^2 +dx_1^2 +dx_2^2+dx_3^2 \right) 
      +\sqrt{f}\left [\frac{dr^2}{\tilde{h}}-\frac{4ml\cosh \alpha}{r^4
      \triangle f}\sin^2\theta dtd\phi 
      \right. \nonumber \\
     &+& \left. r^2 (\triangle d\theta^2 +\tilde{\triangle}\sin^2\theta 
       d\phi^2 +\cos^2\theta d\Omega_3^2\right],
\end{eqnarray}
where 
\begin{eqnarray}
&& f=1+\frac{2m \sinh^2\alpha}{r^4\triangle}, \nonumber \\
&& \triangle =1+\frac{l^2\cos^2\theta}{r^2}, \nonumber\\
&& \tilde{\triangle}=1+\frac{l^2}{r^2} +\frac{2m l^2\sin^2\theta}
         {r^6\triangle f}, \nonumber \\
&& h=1-\frac{2m}{r^4 \triangle}, \nonumber \\
&& \tilde{h}=\frac{1}{\triangle}\left (1 +\frac{l^2}{r^2}-
\frac{2m}{r^4}\right). \nonumber 
\end{eqnarray}
The rotating black three-branes (\ref{e1}) have the horizon at $r_+$
determined by the equation, $\tilde{h}=0$, the horizon is
\begin{equation}
\label{e7}
r_+^2=\frac{1}{2}\left( \sqrt{l^4+ 8m}-l^2\right ).
\end{equation} 
From Eq. (\ref{e1}) we can directly read off the ADM mass and angular 
momentum per unit world volume 
\begin{eqnarray}
&& M=\frac{5\pi ^3 m}{\kappa^2}\left(1+\frac{4}{5}\sinh^2\alpha \right),
      \\
&& J=\frac{2\pi ^3m}{\kappa^2}l \cosh\alpha .
\end{eqnarray}
Here $\kappa^2$ is related to the gravitational constant in ten dimensions
through $\kappa^2 =8 \pi G_{(10)}$. Using the charge quantization condition 
\cite{Kleb}, one may have the number $N$ of the coincident D3-branes
\begin{equation}
N=\frac{4\pi ^{5/2}m}{\kappa}\sinh\alpha \cosh\alpha.
\end{equation}
The angular velocity $\Omega_H$ of the rotating black three-branes is
\begin{equation}
\Omega =\frac{lr_+^2}{2m \cosh\alpha}.
\end{equation}
The Hawking temperature and entropy of the rotating black three branes 
are
\begin{eqnarray}
T  &=& \frac{r_+(2r_+^2 +l^2)}{4\pi m \cosh \alpha}, \\
S  &=& \frac{4\pi^4 mr_+}{\kappa^2} \cosh\alpha.
\end{eqnarray}
In fact, the entropy is the entropy density per unit world volume (throughout 
this paper all extensive quantities are taken as corresponding densities).
These thermodynamic quantities obey the first law of black hole thermodynamics
\begin{equation}
dM=TdS + \Omega dJ + \Phi dN,
\end{equation}
where the chemical potential $\Phi$ corresponding to the number $N$ of the 
D3-branes 
$ \Phi= \pi^{1/2} \kappa ^{-1}\sinh\alpha/ \cosh\alpha. $
Since we are interested in the connection between the low energy excitation 
of the D3-branes and the ${\cal N}$=4 large  $N$ super Yang-Mills theory, 
we have to hold $N$ fixed and to consider the near-extremal configuration 
of the black three-branes. The near-extremal limit may be approached by 
taking $m\rightarrow 0$ and $\alpha \rightarrow \infty$, while $N$ is 
kept as a constant. In this near-extremal limit, all physical quantities
can be expressed in terms of the parameters $m$ and $l$. Using 
$ e^{2\alpha} \approx \frac{\kappa N}{m \pi ^{5/2}}, $
these  related thermodynamic quantities become
\begin{eqnarray}
\label{e12}
&& E=3 \pi ^3 \kappa^{-2} m,  \nonumber \\
&& J= \pi ^{7/4}\kappa^{-3/2}N^{1/2}m^{1/2}l, \nonumber \\
&& \Omega =\pi^{5/4}\kappa^{-1/2}N^{-1/2}m^{-1/2}l r_+^2, \nonumber \\
&& T=2^{-1}\pi^{1/4}\kappa^{-1/2}N^{-1/2}m^{-1/2}(2r^3_+ +l^2r_+),
        \nonumber\\
&&  S=2 \pi^{11/4}\kappa^{-3/2}N^{1/2}m^{1/2}r_+.
\end{eqnarray}
Here $r_+$ is still given by Eq. (\ref{e7}) and $E$ denotes the energy density
of excitations which equals  the ADM mass density of the black three-brane
minus the mass density of the extremal one. It is easy to verify that these 
quantities in Eqs.(\ref{e12}) satisfy the first law of thermodynamics
\begin{equation}
dE=TdS +\Omega  dJ.
\end{equation}
This rotating three-brane has the same quantum numbers as $N$ coincident 
D3-brane with a density of R-charge on the world volume equal to the angular
momentum density $J$. For the D3-brane case, the R-symmetry group is precisely 
the rotating group SO(6) in the dimensions transverse to the brane. Thus the
angular velocity $\Omega$ conjugate to the angular momentum density can be 
called the voltage \cite{Gubser4}. These thermodynamic quantities 
in Eqs. (\ref{e12}) should describe the thermodynamics of the ${\cal N}$=4
large $N$ super Yang-Mills theory on the world volume.

\subsection{Grand canonical ensemble}

Since we keep the spatial  volume of the world volume unchanged, to discuss 
the critical behavior of the conformal field theory on this world volume,
 it is convenient to use a grand canonical ensemble as done in a magnetic 
system \cite{Book,Lousto}. In this ensemble, the appropriate thermodynamic
potential is the Gibbs free energy, which is defined in this case as
\begin{equation}
\label{e18}
G=E-TS -\Omega J,
\end{equation}
and its infinitesimal variation is
\begin{equation}
\label{e19}
dG=-S dT -Jd\Omega.
\end{equation}
In this ensemble, two thermodynamic variables are temperature $T$ and the 
voltage  $\Omega$. Here the temperature $T$ is just the Hawking 
temperature \cite{Gubser2} and the voltage $\Omega$ is the angular velocity
of the black three-brane. 
According to the definition
$ C_{\Omega}=T\left (\frac{\partial S}{\partial T}\right)_{\Omega}$, we
find that  the heat capacity at the constant voltage,
\begin{equation}
\label{e20}
C_{\Omega}=2\pi ^{1/4}\kappa^{-3/2}N^{1/2}
      \frac{\sqrt{m}r_+(2r_+^2+l^2)(3r_+^2-l^2)}
       {(r_+^2 +l^2 )(2r_+^2 -l^2)},
\end{equation}
is positive if $l^2 < 2r_+^2$ or $l^2> 3r_+^2$, and negative between them;
it is divergent at $l^2=2r_+^2$. Note from Eq. (\ref{e7}) that $l^2<2r_+^2$ 
is equivalent to the condition
\begin{equation}
\label{crit1}
l^4/m <8/3,
\end{equation}
which just corresponds to  the stability condition 
 $\chi \equiv \frac{27 \pi^2}{8 N^2}
\frac{J^4}{E^3} <1/3$ found in \cite{Gubser4}, there to derive this condition
Gubser has used the argument that the Hessian matrix of the entropy with 
respect to the energy and angular momentum densities has no positive 
eigenvalues.  According to the duality between
the conformal field theory and supergravity on anti-de Sitter spacetimes
 \cite{Mald}, this indicates that the ${\cal N}$=4  large $N$ 
super Yang-Mills theory  has a phase transition, which is characterized by the 
R-charge. The critical point is at $l^4/m= 8/3$. Beyond this critical point,
the heat capacity is negative. This should be pointed out that
this does not mean that the gauge field enters the negative heat capacity 
phase, but implies that the description via the rotating black three-branes 
breaks down, as happens in the black holes in anti-de Sitter 
spaces \cite{Witten2}.
At this critical point, the critical temperature $T_c$ and critical voltage
$\Omega_c$ are
\begin{eqnarray}
&& T_c=2 \pi ^{1/4} 3^{-1/2} \kappa^{-1/2}N^{-1/2} l, \nonumber \\
&& \Omega_c= 2^{1/2}3^{-1/2}\pi ^{5/4} \kappa ^{-1/2}N^{-1/2}l, 
\end{eqnarray}
which satisfy $\Omega_c/T_c=1/\sqrt{2}\pi $.

Now we are going to investigate the critical behavior of the conformal field 
theory at this critical point. 
Before doing so, it is of some interest
to compare this system with a magnetic system. As is well known, there is 
a continuous phase transition occurring from an ordered
ferromagnetic state to a paramagnetic state in a magnetic system 
\cite{Book,Lousto}. The critical
point is at zero applied magnetic field $H$ and at the critical temperature
$T_c$ the derivative of the magnetization $M$ diverges. In this magnetic 
system, the variation of the Gibbs free energy is \cite{Book}
\begin{equation}
dG=-SdT -MdH.
\end{equation}
Therefore, there is an analogy between our system and the magnetic system:
$M \leftrightarrow  J$, $ H \leftrightarrow \Omega $. Due to this analogy,
one can define some critical exponents related to certain thermodynamic 
quantities. As usual, a critical exponent describing the behavior
near the critical point of a general function $f(x)$ is defined as 
\begin{equation}
\sigma =\lim _{\epsilon \rightarrow 0}\frac{ \ln f(\epsilon)}{\ln\  \epsilon},
      \ ~~~~~~ \ \epsilon =\frac{x-x_c}{x_c},
\end{equation}
where $\sigma $ is called the critical exponent of the function $f(x)$ and
$x_c$ is the critical value of variable $x$. Following the definition 
of critical exponents for a magnetic system \cite{Book,Lousto}, for our system 
we define
\begin{equation}
\label{e24}
C_{\Omega} =T\left (\frac{\partial S}{\partial T}\right)_{\Omega} 
      \sim \left \{
\begin{array}{ll}
     \epsilon _T ^{-\alpha} & \ \ {\rm for}\ \ \epsilon_{\Omega}=0 \\
    \epsilon_{\Omega}^{-\varphi} & \ \ {\rm for}\ \ \epsilon_T=0,
\end{array} \right.
\end{equation}
for the heat capacity at the constant voltage, where 
$\epsilon_T=\frac{T-T_c}{T_c}$ and $\epsilon_{\Omega}= 
\frac{\Omega_c -\Omega }{\Omega_c}$. For the charge density
\begin{equation}
J-J_c \sim \left \{
\begin{array}{ll}
  \epsilon_T ^{\beta} & \ \ {\rm for}\ \epsilon_{\Omega}=0\\
  \epsilon_{\Omega}^{\delta^{-1}} & \ \ {\rm for} \ \epsilon_T=0.
\end{array} \right.
\end{equation}
For the isothermal capacitance 
\begin{equation}
\chi _T =\left(\frac{\partial J}{\partial \Omega}\right)_T
      \sim \left\{
\begin{array}{ll}
  \epsilon_T ^{-\gamma} & \ \ {\rm for} \ \epsilon_{\Omega}=0 \\
 \epsilon_{\Omega}^{\delta ^{-1}-1} & \ \ {\rm for} \ \ \epsilon_T=0.
\end{array} \right.
\end{equation}
And for the entropy 
\begin{equation}
\label{e27}
S-S_c \sim \left \{
\begin{array}{ll}
 \epsilon_T ^{1-\alpha} & \ \ {\rm for}\ \ \epsilon_{\Omega}=0 \\
 \epsilon_{\Omega}^{\psi} & \ \ {\rm for}\ \ \epsilon_T=0.
\end{array} \right.
\end{equation}
   
A  remarkable feature of critical points is the so-called static scaling 
hypothesis \cite{Book}, which asserts that near the critical point, the
singular part of the Gibbs free energy is a generalized homogeneous function.
For our case, we may write
\begin{equation}
G (\lambda ^p \epsilon_T, \lambda^q \epsilon_{\Omega})=\lambda G(\epsilon_T,
   \epsilon_{\Omega}),
\end{equation}
where $p$ and $q$ are two scaling parameters, 
and $\lambda$ is an arbitrary constant.
Thus, those critical exponents defined in Eqs. (\ref{e24})-(\ref{e27}) can be 
expressed  in terms of these two constants $p$ and $q$.  With the help of
Eq. (\ref{e19}), one may arrive at 
\begin{eqnarray}
&& C_{\Omega}=-T\left(\frac{\partial ^2 G}{\partial T^2}\right)_{\Omega},
    \ \
 J=-\left (\frac{\partial G}{\partial \Omega}\right)_T, 
    \nonumber \\
&& \chi_T=-\left(\frac{\partial ^2G}{\partial \Omega^2}\right)_T, 
   \ \  
 S=-\left(\frac{\partial G}{\partial T}\right)_{\Omega}.
\end{eqnarray}
And then one has \cite{Book}
\begin{eqnarray}
&& \alpha =2-\frac{1}{p}, \ \ \beta =\frac{1-q}{p}, \ \ \delta=\frac{q}{1-q},
      \nonumber \\
&& \gamma=\frac{2q-1}{p}, \ \ \psi =\frac{1-p}{q},  
         \ \ \varphi =\frac{2p-1}{q}.\end{eqnarray}
Eliminating $p$ and $q$, one has a set of equalities among those critical 
exponents called scaling laws \cite{Book,Lousto}
\begin{eqnarray}
&& \alpha + 2\beta +\gamma =2, \nonumber \\
&& \alpha +\beta (\delta +1)=2, \nonumber \\
&& \gamma (\delta +1)=(2-\alpha)(\delta -1), \nonumber \\ 
&& \gamma= \beta(\delta -1), \nonumber \\
&& (2-\alpha)(\delta \psi -1)+1 =(1-\alpha)\delta, \nonumber \\
\label{e31}
&& \varphi +2 \psi -\delta ^{-1} =1.
\end{eqnarray}
It should be noticed that these equations are not independent.

Now we calculate the critical exponents to show that they satisfy the 
scaling laws (\ref{e31}). Using  those thermodynamic 
quantities in Eqs. (\ref{e12}), we obtain
\begin{equation}
\chi _T=\kappa^{-1}\pi^{1/2}N mr_+^{-2}
      \frac{2r_+^4 +5l^2r_+^2 -l^4}{2r_+^4 +3l^2r_+^2 +l^4}
     \frac{2r_+^2 +l^2}{2r_+^2 -l^2}.
\end{equation}
According to the definitions (\ref{e24})-(\ref{e27}), we find 
\begin{equation}
\label{e33}
\alpha =\varphi= \beta =\psi = \gamma =\delta ^{-1}=1/2.
\end{equation}
The critical exponent of the heat capacity at the constant
voltage was also obtained in \cite{Gubser4}.  Note from Eq. (\ref{e33}) 
that a  peculiar  feature of these
critical exponents is that they are all equal to $1/2$. However, it is 
easy to verify that they  indeed satisfy the static scaling 
laws (\ref{e31}). 

An important physical quantity describing phase transitions is the 
correlation function, which in general  has  the form \cite{Book,Book1}
\begin{equation}
G(r) \sim \frac{e^{-r/\xi}}{r^{d-2+\eta}},
\end{equation}
where $\xi$ is the correlation length and it diverges at the critical points,
  $\eta$ is the Fisher exponent, and $d$ is the dimensionality of space.
Defining the critical exponent $\nu $ related to the correlation length as
\begin{equation}
\xi \sim \epsilon_T ^{-\nu} \ \ {\rm for}\ \   \epsilon_{\Omega}=0,
\end{equation}
one has a set of equalities related to $\nu$ and $\eta$ \cite{Book}
\begin{eqnarray}
&& d\nu =2-\alpha, \nonumber \\
&& d-2+\eta =\frac{2d}{\delta +1}=\frac{2d\beta}{2-\alpha}
    =\frac{2\beta}{\nu}, \nonumber \\
&& (2-\eta) \nu =\gamma, \nonumber \\
\label{e36}
&& d\frac{\delta-1}{\delta +1}=2-\eta.
\end{eqnarray}
Once again, these equations are not independent, either.
According to the Maldacena's  conjecture \cite{Mald}, the correlation
 function of some conformal field theory can be calculated 
\cite{Gubser1,Witten1}. However, the correlation function for the
 ${\cal N}$=4 super-Yang-Mills theory at finite temperature seems 
 not yet to have been  calculated. Now we are going to 
get the critical exponents $\nu$ and  $\eta $ using the scaling 
laws (\ref{e36}).
For the low energy physics on the D3-brane, its world volume is four 
dimensional, so we have $d=3$. Thus one can get easily from Eqs. (\ref{e36})
\begin{equation}
\nu =1/2,  \ \ \ \eta=1.
\end{equation} 
It would be interesting to verify the critical behavior of the correlation 
function  using the naive field theory model suggested in \cite{Gubser4}.

For the static black three branes, it is known that its thermodynamics is 
always stable, which is indicated by the positive definite heat capacity.
As a consequence, the gauge field described by the supergravity solution is
always in the high temperature  deconfinement phase. That is, the 
Hawking-Page phase transition does not take place in the static D3-branes. 
Because of the above instability of thermodynamics,
it becomes interesting to investigate whether the Hawking-Page
 phase transition appears or not in the rotating D3-branes.
 Recall that the criterion of the Hawking-Page phase transition
  is the sign change of the Euclidean action $I$ of black hole 
solutions \cite{Hawking}.  In the grand canonical ensemble, we have 
$I=T^{-1} G$. From Eqs. (\ref{e12}) we have 
\begin{equation}
G=-\frac{1}{2} \pi ^3 \kappa^{-2}r_+^2 (r_+^2 +l^2),
\end{equation}
from which it is easy to see that the Euclidean action is always 
 negative.  We conclude therefore that the Hawking-Page phase transition 
does not occur in the rotating D3-branes; as in the static D3-branes, the 
system is always in the high temperature deconfinemnet phase, despite 
the instability of thermodynamics.

\subsection{Canonical ensemble}

In \cite{Book} it is proven that if a function with two variables 
$f(x,y)$ is a generalized homogeneous function, then the Lagrendre 
transforms of $f(x,y)$ are also generalized homogeneous functions.
 This result has the immediate consequence 
that  if  one of thermodynamic potentials is a generalized homogeneous 
function near the critical point, other thermodynamic potentials must 
also be generalized homogeneous functions \cite{Book}. For instance, for 
a magnetic system, if the Gibbs free energy has the scaling relation
$$ G(\lambda ^{a_{\epsilon}}\epsilon, \lambda^{a_{H}}H)=
     \lambda G(\epsilon, H), $$
the Helmholtz free energy then must have the scaling relation
$$ F(\lambda^{a_{\epsilon}}\epsilon, \lambda^{a_{M}}M)
            =\lambda F(\epsilon,M),$$
with $a_M=1-a_H$. In fact, this is a consequence of the equivalence of 
thermodynamic ensembles. In the previous subsection we have discussed 
the critical behavior in the grand canonical ensemble. In this subsection 
we do the same thing, but in the canonical ensemble. In this case, the 
appropriate thermodynamic function is the Helmholtz free energy defined as
\begin{equation}
F=E-TS,
\end{equation}
and its variation is
\begin{equation}
dF=-SdT +\Omega dJ.
\end{equation}
In this ensemble the thermodynamic variables are the temperature $T$ and 
the charge density $J$. Other thermodynamic quantities can be expressed 
in terms of them. We note that the heat capacity at the constant charge
\begin{equation}
C_J=\left (\frac{\partial E}{\partial T}\right)_J=
  12 \pi ^{11/4}\kappa^{-3/2}N^{1/2}m^{3/2}r_+^{-1} 
    \frac{2r_+^2 +l^2}{2r_+^4 +5l^2r_+^2 -l^4},
\end{equation}
is positive when 
\begin{equation}
l^4/m < (l^4/m)_c=  \frac{19 +\sqrt{297}}{4}\approx 9.058;
\end{equation}
 negative as $l^4/m >(l^4/m)_c$; and diverges at $(l^4/m)_c$. Comparing 
with Eq. (\ref{crit1}), it is interesting to observe that these two critical 
points are not in agreement with each another. At $l^4/m=8/3$, no special
happens in the heat capacity $C_J$. Also we can check that no first 
derivatives and second  derivatives of the Helmholtz free energy 
diverge at $l^4/m =8/3$. This implies that in this ensemble $l^4/m=8/3$
is not a critical point. Instead $l^4/m =9.058$ is a critical point 
according to  the definition of critical points: one of second derivatives
of certain thermodynamic potential diverges there. Indeed two physical 
quantities $C_J$ and $\chi ^{-1}_T$, which are second derivatives of the 
Helmholtz free energy with respect to $T$ and $J$, respectively, diverge 
at $l^4/m =(l^4/m)_c$. At this critical point, the critical 
temperature is
\begin{equation}
 T_c =2^{-3/2} \pi ^{1/4} \kappa^{-1/2} N^{-1/2} l x^{-1/4} (8+x)^{1/2}
     ((8+x)^{1/2}-x^{1/2})^{1/2},
\end{equation}
where $x=(19+\sqrt{297})/4$. 
Using corresponding definitions of critical exponents, that is, in 
Eqs. (\ref{e24})-(\ref{e27}) replacing $C_{\Omega}$ by $C_J$, $\chi _T$ by 
$\chi^{-1}_J$, $J$ by $\Omega$, and $\epsilon _{\Omega}$ by $\epsilon_J$, 
we find  that these critical exponents have  the same values as those 
in Eq. (\ref{e33}), i.e., 
\begin{equation}
 \alpha =\varphi =\beta =\psi =\gamma =\delta ^{-1}=1/2,
\end{equation}
and then they also  satisfy the same scaling laws (\ref{e31}).

This interesting result gives rise to the question whether these 
two ensemble are equivalent or not. From our results it indicates 
that these two ensembles are not equivalent to one another.
 The critical point $l^4/m =8/3 \approx 2.667$ in the grand 
canonical ensemble is shifted to $l^4/m \approx 9.058$ in the canonical 
ensemble.  Indeed there have been some arguments that 
thermodynamic ensembles are not equivalent in the self-gravitating 
system \cite{Parentani,Miller}. And the thermodynamic stability depends on
the chosen environment (ensemble) \cite{Okamoto}. Therefore, the stability
boundary in \cite{Gubser4} is given in the grand canonical ensemble. 
In the canonical ensemble the stability boundary is $l^4/m <9.058$.
In both of two  critical points, corresponding critical exponents are 
same and satisfy the static scaling laws. Furthermore, 
in the canonical ensemble, the Euclidean action is $T^{-1}F$,
 from Eqs. (\ref{e12}) we have 
\begin{equation}
F=-\frac{1}{2}\pi^3 \kappa^{-2}r_+^2(r_+^2 -l^2).
\end{equation}
This free energy is negative if $r_+^2>l^2$, that is $l^4/m<1$. But it becomes
positive as $l^4/m>1$ and changes its sign at $l^4/m=1$. This indicates that
the Hawking-Page phase transition may occur in the canonical ensemble.

\section{Critical behavior in rotating M5-branes}

In this section we discuss the critical behavior for the rotating M5-branes
in M theory. Similar to the previous section, we also consider  
the case in which an angular momentum parameter does not vanish only.  The 
rotating M5-brane solution has been found by Cvetic and Youm in \cite{Cvetic}.
The metric  can be written down as
\begin{eqnarray}
\label{3e1}
ds^2_{11} &=& f^{-\frac{1}{3}}(-hdt^2 +dx_1^2 +\cdots +dx_5^2) + 
       f^{\frac{2}{3}}\left [\frac{dr^2}{\tilde{h}} + r^2 (\triangle 
       d\theta^2 +\tilde{\triangle}\sin^2 \theta d\phi^2 
       \right. \nonumber \\
  &+& \left. \cos^2\theta d\Omega^2_2)
    -\frac{4ml \cosh\alpha}{r^3\triangle f}\sin^2\theta dtd\phi \right],
\end{eqnarray} 
where
\begin{eqnarray}
&& f=1+\frac{2m \sinh^2\alpha}{r^3\triangle}, \nonumber \\
&& \triangle =1+\frac{l^2\cos^2\theta}{r^2}, \nonumber \\
&& \tilde{\triangle}=1+\frac{l^2}{r^2} +\frac{2m l^2
       \sin^2\theta}{r^5\triangle f}, \nonumber \\
&& h=1-\frac{2m}{r^3\triangle}, \nonumber\\
&& \tilde{h}=\frac{1}{\triangle}\left[1+\frac{l^2}{r^2}-\frac{2m}{r^3}\right].
    \nonumber
\end{eqnarray}
Through a double dimensional reduction, the rotating black M5-brane solution 
can be  reduced to a rotating black D4-brane solution in the
 tpye IIA supergravity.  In the reduction, thermodynamics is kept unchanged. 
The rotating black M5-brane (\ref{3e1})
 has the horizon determined by the positive real root of the equation
\begin{equation}
\label{3e5}
r_+^3 +l^2r_+ -2m =0.
\end{equation}
Those thermodynamic quantities relevant to the rotating black 
M5-branes are readily obtained. The result is 
\begin{eqnarray}
&& M=\frac{32 \pi^2 m}{3\kappa^2} \left (1+\frac{3}{4}
   \sinh^2\alpha\right), \\
&& J=\frac{16 \pi^2 ml}{3\kappa^2}\cosh \alpha, \\
&& \Omega= \frac{lr_+}{2m \cosh\alpha}, \\
&& T=\frac{3r_+^2 +l^2}{8\pi m \cosh\alpha}, \\
&& S=\frac{32\pi ^3mr_+}{3\kappa^2}\cosh\alpha.
\end{eqnarray}
where $\kappa^2=8\pi G_{(11)}$ related to the gravitational constant in 
eleven dimensions. 
According to the charge quantization condition \cite{Kleb}, the number $N$ 
of the coincident M5-branes is 
\begin{equation}
N=2^{10/3}\pi^ {5/3} \kappa^{-2/3} m \cosh \alpha \sinh\alpha.
\end{equation} 
These quantities trivially satisfy the first law of black hole thermodynamics
$$ dM = TdS +\Omega dJ + \Phi dN,$$
where the chemical potential 
$ \Phi = 2^{-1/3} \pi ^{-4/3} \kappa^{-2/3} \sinh\alpha/ \cosh \alpha.$
As in the case of D3-branes, to connect the thermodynamics of black 
M5-branes and that of the conformal field theory on the world volume, we take 
the near- extremal limit.  Taking $m\rightarrow 0$ and $\alpha \rightarrow 
\infty$ and keeping $N$ fixed, we obtain
\begin{eqnarray}
&& E=20 \pi^2 3^{-1}  \kappa^{-2} m, \nonumber \\
&& J=2^{7/3} 3^{-1}\pi ^{7/6} \kappa^{-5/3} N^{1/2}m^{1/2}l, \nonumber \\
&& \Omega =2^{2/3}\pi^{5/6}\kappa^{-1/3}N^{-1/2}m^{-1/2}lr_+, \nonumber \\
&& T=2^{-4/3}\pi ^{-1/6}\kappa^{-1/3}N^{-1/2}m^{-1/2}(3r_+^2+l^2),
          \nonumber\\
&& S=2^{10/3} 3^{-1} \pi ^{13/6}\kappa^{-5/3}N^{1/2}m^{1/2}r_+.
\end{eqnarray}
We first consider the thermodynamic stability
of rotating M5-branes in grand canonical ensemble. The heat capacity
at the constant angular velocity (voltage) is
\begin{equation}
\label{3e17}
C_{\Omega}=2^{10/3} 3^{-1}\pi ^{13/6}\kappa^{-5/3} N^{1/2}m^{1/2}r_+
     \frac{15r_+^4 +3l^2r_+^2 -2l^4}{3r_+^4 +4l^2r_+^2 +l^4}
      \frac{3r_+^2 +l^2}{3r_+^2-l^2}.
\end{equation}
This heat capacity is positive  when
\begin{equation}
\label{3e18}
l^2 <3r_+^2, \ \ {\rm that\ \ is,}\ \ l^3/m < \sqrt{27} /2
\end{equation}
or $2l^4 >15r_+^4 +3l^2 r_+^2$, negative between them, and diverges 
at  $l^2=3r_+^2$. Therefore the thermodynamically stable boundary is
$l^3/m <\sqrt{27}/2$ in this ensemble. At the critical point $l^2=3r_+^2$,
the isothermal capacitance  also diverges  as
\begin{equation}
\chi_T =\left(\frac{\partial J}{\partial \Omega}\right)_T
  =2^{5/3} \pi^{1/3} \kappa^{-4/3} Nmr_+^{-1} 
   \frac{r_+^4 +4l^2r_+^2 -l^4}{3r_+^4 +4l^2r_+^2 +l^4}
  \frac{3r_+^2 +l^2}{3r_+^2-l^2}.
\end{equation}
The critical temperature $T_c$ and critical voltage $\Omega_c$ are
\begin{eqnarray}
&& T_c=2^{-5/6} 3^{3/4} \pi ^{-1/6} \kappa^{-1/3} N^{-1/2}l^{1/2},
    \nonumber \\
&& \Omega_c= 2^{-7/6} 3^{1/4} \pi ^{5/6} \kappa^{-1/3} N^{1/2}l^{1/2},
\end{eqnarray}
they obey $\Omega_c/T_c =2\pi/\sqrt{3}$.
Furthermore, one can easily find that those critical exponents defined 
in Eqs. (\ref{e24})-(\ref{e27}) in this case have the same value as those 
in Eq. (\ref{e33}). However,
due to the fact that the low energy excitation of M5-branes should be  
described by  a six dimensional (0,2) conformal field theory in  
the large $N$ limit \cite{Mald}, the critical exponents related to the 
correlation function will be changed. Considering $d=5$, we find from 
Eqs. (\ref{e36})
\begin{equation}
\nu =3/10, \ \ \  \eta=1/3.
\end{equation}
 
We now turn to the canonical ensemble. In this ensemble, some second 
derivatives of the Helmholtz free energy are characterized by the heat
capacity at constant angular momentum (R-charge) and the inverse
isothermal capacitance  $\chi_T ^{-1}$. 
Note that both  the heat capacity
\begin{equation}
C_J=2^{13/3}3^{-2}5 \pi ^{13/6}\kappa^{-5/3}N^{1/2} m^{3/2}
    \frac{3r_+^2 +l^2}{r_+^4 +4l^2r_+^2 -l^4},
\end{equation}
and $\chi_T^{-1}$ diverge at
\begin{equation}
\label{3e22}
l^4=r_+^4 +4l^2 r_+^2.
\end{equation}
Obviously, this critical point is different from that in Eq. (\ref{3e18}).
That is, the critical point is shifted because of the different ensemble.
However, once again, it is trivial to verify that at the critical 
point (\ref{3e22}) corresponding critical exponents have  the same  values 
as those at the critical point (\ref{3e18}). In addition, we find
\begin{eqnarray}
&& G=-\frac{2}{3}\pi ^2 \kappa^{-2} (r_+^3 +l^2r_+), \\
&& F=-\frac{2}{3}\pi ^2 \kappa^{-2}(r_+^3 -3l^2r_+).
\end{eqnarray}
Because the Gibbs free energy is always negative, the Hawking-Page transition 
will not appear in the grand canonical ensemble; the conformal field i
s always in the high temperature phase.
But the Helmholtz free energy will change its sign at $r_+^2=3l^2$. And hence
the Hawking-Page transition may occur in the canonical ensemble for the
rotating M5-branes.

\section{Critical behavior in rotating M2-branes}

In this section we consider the case of the rotating M2-branes. 
The rotating black M2-brane metric with a nonvanishing angular momentum 
is \cite{Cvetic}
\begin{eqnarray}
\label{4e1}
ds_{11}^2 &=& f^{-2/3} (-hdt^2 +dx^2_1+dx^2_2) + f^{1/3}\left [ \frac{dr^2}
     {\tilde{h}} +r^2 (\triangle d\theta^2 +\tilde{\triangle}
     \sin^2\theta d\phi^2  \right. \nonumber \\
   &+& \left. \cos^2\theta d\Omega^2_5 )- \frac{4ml \cosh\alpha}
     {r^6\triangle f} \sin^2\theta dtd\phi \right],
\end{eqnarray}
where 
\begin{eqnarray}
&& f= 1+\frac{2m\sinh^2\alpha}{r^6\triangle}, \nonumber \\
&& \triangle = 1+\frac{l^2\cos^2\theta}{r^2}, \nonumber \\
&& \tilde{\triangle}= 1+\frac{l^2}{r^2} +\frac{2ml^2\sin^2\theta}
       {r^8 \triangle f}, \nonumber \\
&& h=1-\frac{2m}{r^6\triangle}, \nonumber \\
&& \tilde{h}= \frac{1}{\triangle}\left (1+\frac{l^2}{r^2}-
        \frac{2m}{r^6}\right). \nonumber 
\end{eqnarray}
The horizon of the rotating black M2-brane (\ref{4e1}) is determined by 
the equation
\begin{equation}
r_+^6 +l^2r_+^4 -2m =0.
\end{equation}
Through a straightforward calculation, we have 
\begin{eqnarray}
&& M= \frac{7\pi^4 m}{3\kappa^2} \left (1+\frac{6}{7}\sinh^2 \alpha \right),\\
&& J=\frac{2\pi^4 ml}{3\kappa^2}\cosh\alpha, \\
&& \Omega = \frac{lr_+^4}{2m \cosh\alpha}, \\
&& T=\frac{r_+^3(3r_+^2 +2l^2)}{4\pi m \cosh \alpha}, \\
&& S= \frac{4\pi^5 mr_+}{3\kappa^2}\cosh\alpha, \\
&& N= 2^{2/3}\pi^{10/3}\kappa^{-4/3}m \cosh\alpha \sinh\alpha.
\end{eqnarray}
which obey the first law of black hole thermodynamics
$$ dM =TdS +\Omega dJ + \Phi dN $$
with $ \Phi =2^{1/3}\pi ^{2/3}\kappa^{-2/3}\sinh\alpha/\cosh\alpha.$
Here  $\kappa^2 $ is related to the  gravitational constant in eleven
 dimensions as in the 
preceding section.  The thermodynamics of the low energy excitations of
the rotating M2-branes can also be obtained by taking the near-extremal limit: 
$ m \rightarrow 0$  and $\alpha \rightarrow \infty $, while keeping $N$ fixed.
 In the limit, we reach  
\begin{eqnarray}
\label{4e12}
&& E=3^{-1} 4 \pi ^4  \kappa^{-2} m, \nonumber \\
&& J= 2^{2/3} 3^{-1}\pi ^{7/3} \kappa^{-4/3}N^{1/2}m^{1/2}l, \nonumber \\
&& \Omega = 2^{-2/3}\pi ^{5/3}\kappa^{-2/3}N^{-1/2}m^{-1/2}lr_+^4,
            \nonumber \\
&& T= 2^{-5/3}\pi ^{2/3}\kappa^{-2/3}N^{-1/2}m^{-1/2}r_+^3(3r_+^2 +2l^2), 
      \nonumber \\
&& S = 2^{5/3}3^{-1}\pi^{10/3}\kappa^{-4/3}N^{1/2}m^{1/2}r_+.
\end{eqnarray}

For the  rotating M2-branes, from Eqs.(\ref{4e12}) we have  
\begin{equation}
\label{4e17}
C_{\Omega}=2^{5/3}3^{-1}\pi ^{10/3}\kappa^{-4/3}N^{1/2}m^{1/2}r_+
    \frac{6r_+^4 +l^2r_+^2 -2l^4}{3r_+^4 +5l^2r_+^2 +2 l^4}
    \frac{3r_+^2 +2l^2}{3r_+^2 -2l^2}.
\end{equation}
It is easy to see that this heat capacity is positive 
as 
\begin{equation}
2l^2 <3r_+^2,\ \  {\rm that \ \ is, }\ \ l^6/m <27/10,
\end{equation}
or $2l^4 >6r_+^4 +l^2r_+^2 $, negative between them, and diverges 
at $2l^2=3r_+^2$. At this critical point, the isothermal capacitance  
diverges as well:
\begin{equation}
\label{4e19}
\chi_T= 2^{4/3}\pi ^{2/3} \kappa^{-2/3}N m r_+^{-2}
         \frac{r_+^2 +2l^2}{ 3r_+^4 +5l^2r_+^2 +2l^4}
         \frac{3r_+^2 +2l^2}{3r_+^2 -2l^2}.
\end{equation}
In this case, the critical temperature and critical voltage
are
\begin{eqnarray}
&& T_c= 2^{4/3} 5^{-1/2}\pi ^{2/3} \kappa ^{-2/3} N^{-1/2} l^2, 
   \nonumber \\
&& \Omega_c = 2^{5/6} 3^{-1/2}5^{-1/2}\pi ^{5/3}
      \kappa ^{-2/3}N^{-1/2} l^2 , 
\end{eqnarray}
that is, $\Omega_c/T_c=\pi/\sqrt{6} $.

As for the critical exponents at this critical point, 
once again, we find those critical exponents defined  in 
Eqs. (\ref{e24})-(\ref{e27}) are the same as those (\ref{e33}), which seems to 
show the universality of this kind of critical behaviors in the 
conformal field theory.  Note that the low energy physics on the M2-brane 
should be described by
a conformal field theory in three dimensions \cite{Mald,Kleb}. Thus $d=2$
in this case,  from  Eqs. (\ref{e36}), we  furthermore have the critical 
exponents of correlation function
\begin{equation}
\nu= 3/4, \ \ \ \eta= 4/3.
\end{equation}
Differing from the cases of D3-branes and M5-branes, we find that the 
heat capacity at constant angular momentum (R-charge) for the M2-branes 
\begin{equation}
C_J=  2^{11/3} 3^{-2}   \pi ^{10/3} \kappa^{-4/3} N^{1/2}m^{3/2} r_+^{-5}
    \frac{3r_+^2+2l^2}{r_+^2 +2l^2},
\end{equation}
is always positive. Also the inverse isothermal  capacitance  
has not any divergent point, which can be see clearly from Eq. (\ref{4e19}).
This means that the Helmholtz free energy has not any singular point,
at least till its  second derivatives. Therefore in the canonical 
ensemble the  M2-brane is thermodynamically stable and there is no critical
point. In this sense,  it would be interesting to note that there  
exist  phase 
transitions  in the self-gravitating system considered in \cite{Miller} 
in the microcanonical ensemble and conical ensemble, but not in the grand
 canonical ensemble, which also supports the point of view of the 
inequivalence of thermodynamic ensembles.  As for the Hawking-Page phase 
transition, in this case,  we have from Eqs. (\ref{4e12})
\begin{eqnarray}
&& G=-\frac{1}{3} \pi ^4 \kappa^{-2} (r_+^6 +l^2r_+^4), \\
&& F=-\frac{1}{3}\pi ^4 \kappa^{-2} r_+^6 . 
\end{eqnarray}
They are always negative and therefore the Hawking-Page phase transition 
does not occur in both of the grand canonical ensemble and canonical 
ensemble. It is worthwhile to note the difference between the thermodynamics
of rotating D3- and  M5-branes and of the rotating M2-branes.

\section{Conclusions}

 In this work we have investigated the thermodynamic stability, critical 
behavior near the stability boundary, and the Hawking-Page transition 
 for  the low energy excitations of rotating D3-branes, M5-branes and 
M2-branes in the grand canonical ensemble and canonical ensemble,
 respectively. In the superconformal field theory which characterizes 
the low energy excitations of the branes, the angular momentum
is interpreted as electric charge under a subgroup of the R-symmetry group.
Therefore the existence of the stability boundary characterized by the angular
momentum implies the existence of phase transition of the superconformal
field theory, characterized by the R-charge.

The rotating black D3-brane is thermodynamically stable up to a critical
angular momentum density. This indicates that there is, according to the 
Maldacena's conjecture, a corresponding phase transition for 
 the ${\cal N}$=4  large $N$ super Yang-Mills theory at finite temperature,
which is characterized by the R-charge. We have studied this phase transition
on the side of supergravity by calculating some related critical exponents.
Although these critical exponents (\ref{e33}) are all  equal to $1/2$, indeed  
they  are shown to satisfy the static scaling laws.
Using the scaling laws related to the correlation function, we have 
also deduced the critical exponents $\nu=1/2$  and $\eta =1$ of the correlation
function of the gauge field. Of interest we found is that the critical point 
is different in the grand canonical ensemble and canonical 
ensemble, although the corresponding critical exponents at both of these
 two critical points  have the same values and satisfy the same 
static scaling laws. This result seems to imply that indeed thermodynamic 
ensembles concerning black holes are not equivalent 
and thermodynamic stability depends on the chosen ensemble 
\cite{Parentani,Okamoto}.

The rotating  M5-brane and M2-brane have been found to have similar
stability boundary determined by a critical angular momentum value. 
This implies that there is also a phase transition for the corresponding 
conformal field theory \cite{Mald,Kleb}. Some critical exponents
are the same as those for the rotating three branes, which shows the 
universality of this kind of phase transition for the rotating branes.
 But the critical exponents related to the correlation function is different 
due to the different dimensionality of the world volume.

Through the calculation of the Euclidean action of the rotating black 
 branes, we have found that the Hawking-Page transition does not occur 
in the grand canonical ensemble, as in the static D-branes. In the canonical
ensemble, however, the Helmholtz free energy may changes its sign for 
the rotating D3- and M5-branes, but not for the M2-branes. This seemingly
implies that the Hawking-Page phase transition may occur in this ensemble.
In addition, it seems to be
worth noting that in the canonical ensemble  the Helmholtz free energy for 
the rotating  M2-branes is regular and there is no any critical point. 
 Another point is that heat capacity at constant angular
velocity (voltage) is also positive if the  angular momentum density
is large enough [see (\ref{e20}), (\ref{3e17}) and (\ref{4e17})]. 
We have not yet understood its meanings and its relation, if any, to the 
thermodynamic stability.  It would be interesting to understand the shift 
of critical point in  the different ensembles in the field theory 
model suggested in \cite{Gubser4}. Also it would be of some significance to
extend the Gubser's method \cite{Gubser4} to explain the entropy and 
stability boundary for the rotating black M5-branes and M2-branes. Moreover,
 it should be important to check whether the critical exponents of
 correlation function in the field theory model is in agreement with the
 values given in this paper or not.

{\bf Note Added.} Since this paper was finished, there have been 
three related papers \cite{paper1,paper2,paper3} appearing in the 
hep-th archive. Refs.\cite{paper1,paper2}  relate the
rotating D3-, M2-, M5-branes to the charged AdS black hole solutions in 
gauged supergravity theories, and study  some thermodynamic properties 
of the latter.  Ref. \cite{paper3} discusses the thermodynamic stability
 and  the corresponding QCD model for rotating D3-, M5-, and M2-branes 
with multiple angular momentum parameters. An interesting result is that
in the canonical ensemble there may exist critical points for the rotating
M2-branes with multiple angular momentum parameters. As a result, 
we suspect that the Hawking-Page transition may also appear in the
 canonical ensemble of this system.

\section*{Acknowledgments}
This research was supported by the Center for Theoretical Physics
of  Seoul National University. R. G. Cai would like to thank 
Drs. C. Liu and M. I. Park for useful discussions.

\end{document}